\documentstyle[12pt,epsf,bezier]{article}
\begin{document}

\title{
\vspace{-2.0cm}
\begin{flushright}
{\normalsize UTHEP-318}\\
\vspace{-0.3cm}
{\normalsize August 1995}\\
\end{flushright}
\vspace*{3.0cm}
{\Large On the phase structure of QCD
        with Wilson fermions\footnote{
talk presented at ``QCD on massively parallel computers''
held at Yamagata, Japan, on March 16-18, 1995.}
 \vspace*{0.5cm}}
}
\author{S. Aoki\\ \\
{\it Institute of Physics, University of Tsukuba,}\\
{\it      Tsukuba, Ibaraki-305, Japan}
}
\date{}
\maketitle

\vspace*{1cm}
\begin{abstract}
The phase structure of lattice QCD with Wilson
fermions is discussed.
Analytic and numerical evidences are given for
the spontaneous breaking of parity and flavor symmetry, which
naturally explains the existence of the massless pion at the critical
hopping parameter $K_c$ without recourse to
the chiral symmetry absent in the Wilson fermion formulation.
New numerical evidences are also presented
for the multiple structure of the critical lines
in the weak coupling region.
A connection between the phase structure and the finite temperature
phase transition is briefly mentioned.
\end{abstract}

\newpage

\section{Introduction}

Elucidating the nature of chiral transition separating the high-temperature
quark-gluon plasma phase from the low-temperature hadron phase has
been one of the focal points of effort in recent numerical
simulations of full lattice QCD including light dynamical quarks.  Much of work
in this direction has employed the Kogut-Susskind quark
action since it retains a $U(1)$ subgroup of chiral symmetry.  On the other
hand, studies with the Wilson quark action are less well
developed in spite of the effort over the
years\cite{earlywork,qcdpaxtwo,milc}.

A main problem for studies with Wilson fermions
originates from the presence of explicit chiral symmetry breaking
in the action at non-zero lattice spacing,
which is introduced to avoid the well-known species doubling problem.
Clearly, before serious studies on the phase structure
at finite temperature are initiated,
more understanding on the effect of the explicit chiral symmetry
breaking at zero temperature is necessary.

Let me consider a phase space of $\beta=6/g^2$ and $K\equiv
1/2M = 1/(2m_qa + 8)$
at zero temperature, where $g$ is the bare gauge coupling,
$m_q$ the bare quark mass, and $a$ the lattice spacing.
There is much evidence from
analytical considerations and numerical simulations of
hadron masses that the pion mass vanishes along a line $K=K_c(\beta)$, the
critical line, which runs from $K_c(\beta=0)=1/4$ in the strong-coupling limit
to $K_c(\beta=\infty)=1/8$ in the weak-coupling limit.
However the massless pion appeared on $K_c(\beta)$ can not be
regarded as the Nambu-Goldstone boson of the spontaneous breaking of
chiral symmetry since it is explicitly broken in the Wilson fermion formulation
at non-zero lattice spacing.

Why does the massless pion appear then ? It has conventionally been explained
as follows:
The appearance of the massless pion on $K_c$ may be an evidence
that the tuning $K$ effectively recovers the chiral symmetry,
explicitly broken by the Wilson fermion formulation.
Off course, this in NOT an explanation, since it does not tell
why the massless pion appear on $K_c$.
It also can not tell what happens beyond $K_c$.

Instead of the conventional explanation,
a possible explanation was proposed some years ago, to
explain the appearance of the massless pion\cite{aoki1},
on the basis  of new phase structure of lattice QCD with
Wilson fermions.
Later some analytic\cite{aoki2,aoki3} and
numerical\cite{AG1,AG2,AG3} evidences were
given for the existence of the phase structure,
but it is far from being completely established.
In this report, after reviewing the previous accomplishment,
more evidences are supplied to established the proposed
phase structure.

\section{Possible explanation}

In this section we review the possible explanation for the massless pion
appeared on $K_c$ in the Wilson fermion formulation\cite{aoki1,aoki3}.
Let me consider lattice QCD with a single flavor first.
In the theory there exists one ``$\pi$ meson'',
corresponding to a pseudo-scalar operator
$ \pi (x,t) = \bar\psi i\gamma_5\psi(x,t) $.
At $K < K_c$
the correlation function of the operator for large $t$ behaves
such that
\begin{equation}
\sum_x \langle \pi (0,0) \pi (x,t) \rangle \simeq Z \exp ( - t/\xi)
\end{equation}
where the correlation length $\xi$ is related to the mass of the $\pi$ meson
as $\xi = 1/ (m_\pi a)$.
Since the correlation length diverges on $K_c$
the correlation function for large $t$
behaves at $K=K_c$ such that
\begin{equation}
\sum_x \langle \pi (0,0) \pi (x,t) \rangle \simeq Z t^{-\alpha}
\end{equation}
with some exponent $\alpha$. If we go beyond $K_c$ ($K > K_c$),
it is expected for large $t$ that
\begin{equation}
\sum_x \langle \pi (0,0) \pi (x,t) \rangle \simeq V \langle \pi \rangle^2
+ Z \exp ( - t/\xi)
\end{equation}
with $ \langle \pi \rangle \not=0$ where $V$ is a space volume.

The above situation is summarized as follows.
Since the massless $\pi$ meson implies the divergent correlation length
on $K_c$, the operator corresponding to the $\pi$ meson
has the long-range order at $K > K_c$ in such a way that
$ \langle \pi \rangle \not=0$.
This condensation spontaneously breaks
the discrete parity at $K>K_c$ and the second order
phase transition from the parity symmetric phase to
the parity broken one occurs at $K = K_c$.
Therefore the massless $\pi$ meson can be interpretated as the divergent
correlation length associated with this second order phase transition,
not a Nambu-Goldstone boson.

According to the above interpretation we expect near $K_c$,
$
(m_\pi a)^2 \sim (K_c-K)^{2\nu} \propto (M-M_c)^{2\nu}
$
with another critical exponent $\nu$.
If we regards the effective theory
for the $\pi$ meson as a four-dimensional scalar field theory, we can expect
the phase transition to be mean-field like up to the logarithmic corrections
and therefore $\nu =1/2$, which reproduces well-known PCAC relation that
$
(m_\pi a)^2 \propto m_qa
$
where a quark mass  $m_q a \simeq M-M_c$.

Now we consider QCD with 2 flavors, where
members of the pseudo-scalar mesons are
$\pi^a = \bar\psi i\gamma_5\tau^a\psi$ (non-singlet) and
$\eta =\bar\psi i\gamma_5 \psi$ (singlet).
Since the observed mass spectra of these mesons satisfy
$ m_\eta > m_\pi $,
it is expected at $K = K_c$  that
$
m_\pi a = 0,  m_\eta a \not= 0
$ .
To satisfy this,
there exists a phase transition at $K=K_c$,
which implies the condensation such that
$
\langle \pi^3 \rangle \not=0, \langle \eta \rangle = 0
$
at $K > K_c$.
The above condensation spontaneously breaks
both parity and flavor symmetry. In this case the
neutral $\pi^0$ meson corresponding to the $\pi^3$ operator
is identified with the massless mode of this phase transition and
becomes massless only at the transition point $K=K_c$.
The charged $\pi^\pm$ mesons corresponding to the linear combination of
$\pi^1$ and $\pi^2$ operators are Nambu-Goldstone bosons associated with
the flavor symmetry breaking: $m_{\pi^0} = m_{\pi^\pm}$ in the phase
with $\langle \pi^3\rangle = 0$ and $m_{\pi^\pm} = 0$ in the phase with
$\langle \pi^3\rangle \not= 0$. On the other hand the $\eta$ meson stays
massive for all $K$ since $\langle \eta \rangle $ is always zero.
This phase transition
can naturally solve not only the problem of the massless pion but
also the lattice U(1) problem\cite{aoki3}.

One can rigorously prove that
$\langle \eta \rangle = 0$ using {\it a la} Vafa-Witten theorem\cite{VW},
while no such restriction can be applied to the operator
$\pi^3$, since the Wilson fermion determinant is shown to be
real-positive for $N_f=2$
even in the presence of the source term $ iH\bar\psi \gamma_5\tau^3\psi $.

\section{Evidences for the phase transition}

\subsection{Analytic results}

In the strong coupling ($\beta =0$) and large $N_c$ limit
we can analytically calculated the behavior of
order parameter and meson masses\cite{aoki1,aoki2}:
\begin{eqnarray}
\langle \bar\psi i\gamma_5 \psi \rangle & =  0 &\qquad  M^2 > M_c^2
\nonumber \\
\langle \bar\psi i\gamma_5 \psi \rangle & =\displaystyle  \pm\frac{2}{16-M^2}
\sqrt{3(4-M^2)}\times 4N_c  &\qquad M^2 < M_c^2 \nonumber
\end{eqnarray}
where the critical point is given by $M_c =\pm 2 $($K_c=\pm1/4$)
with $M=1/2K$.
In this limit there is no essential difference between $N_f=1$ and $N_f=2$.
It was also shown that the $\pi$ meson becomes massless
on $M_c$\cite{aoki1,aoki2}.

In the strong coupling and 1/$N_c$ expansions
the behavior of the order parameter
is essentially the same for $N_f=1$ case:
$\langle \bar\psi i\gamma_5 \psi \rangle \not= 0$ at $M^2 < M_c^2(\beta)$, and
the pion becomes massless on $M_c(\beta)$, which was calculated
in the expansion of $\beta$\cite{aoki2,aoki3}.
For $N_f =2$ we obtain
$\langle \bar\psi i\gamma_5\tau^3  \psi \rangle \not= 0$
at $M^2 < M_c^2(\beta)$
while $\langle \bar\psi i\gamma_5 \psi \rangle = 0$ for all $M$,
as expected\cite{aoki2,aoki3}.
Meson spectrum also agrees with our expectation in the previous section.

\subsection{Results from numerical simulations}

To calculate quark propagators on a given configuration
iterative methods such as Conjugate Gradiant (CG) method
are usually used to invert the lattice fermion matrix.
For the case of the Wilson fermion, if $K$ approaches $K_c$,
the inversion becomes more and more difficult:
a number of CG iteration rapidly increases.
To overcome this difficulty and to calculate non-zero order parameter,
we add an external field coupled to the order parameter
and take $H\rightarrow 0$ limit after the average over configurations
is taken.

Let me consider the quenched QCD first.
If we add the external field $ i H \bar\psi \gamma_5 \psi$ to
the Wilson fermion action, it is easy to show that
\begin{equation}
{\rm Re} \langle \bar\psi i\gamma_5 \psi \rangle
(U, H)  =
-{\rm Re} \langle \bar\psi i\gamma_5 \psi \rangle
(U, -H)
\end{equation}
where $\langle O \rangle (U, H)$ means a value of an operator $O$
on a given gauge configuration $U$ and the external source $H$.
If $H\rightarrow 0$ limit was non-singular,
this would lead to
$
{\rm Re} \langle \bar\psi i\gamma_5 \psi \rangle (U, 0) = 0
$
before the average over gauge configurations.
Therefore
$
\langle \bar\psi i\gamma_5 \psi \rangle (H=0 ) = 0
$
in the numerical simulation on a finite lattice,
where the limit is always non-singular.
However this does not means $\langle \bar\psi i\gamma_5 \psi \rangle =0$.
Since, in the parity broken phase,
$H\rightarrow 0$ limit is singular as the volume $V$ goes to infinity
one should take limits in the following order:
\begin{equation}
\lim_{H\rightarrow 0} [ \lim_{V\rightarrow\infty}
\langle \bar\psi i\gamma_5 \psi \rangle (H) ] .
\end{equation}
Since actual numerical simulations are always performed on finite volumes,
one should extrapolate to $H = 0$ from non-zero $H$,
not put $H=0$ in the simulation, in order to
obtain the correct value of $\langle \bar\psi i\gamma_5 \psi \rangle$
at $H=0$ in the infinite volume limit.

This external source method
has been applied to show that
the parity breaking phase transition exists at $T=0$, where
$T$ is temperature,
using the quenched QCD simulation at $\beta =5.0$ on a $4^4$
lattice\cite{AG1}.
Such a phase transition disappear at $T > T_c$,
where $T_c$ is the critical temperature of the pure gauge theory,
using the  quenched QCD simulation at $\beta =5.6$, 5.8 on a
$8^3 \times 4$ lattice\cite{AG2}.

For the case of full QCD with 2 flavors, a term
$ i H \bar \psi i\gamma_5 \tau^3 \psi$ should be added.
Even in the presence of the term the Wilson fermion determinant
is still real-positive. Therefore the hybrid Monte Carlo method
is applicable to this case. Again we have to keep $H$ non-zero
for simulations on a finite volume, in order to find
the spontaneous breakdown of parity and flavor symmetry.

The hybrid Monte Carlo simulation has been performed
for the full QCD with $N_f=2$ at $\beta =5.0$
on $4^4$, $4^2\times 6^2$, and $6^4$ lattices\cite{AG3}.
Although some evidence for the parity and flavor symmetry breaking phase
transition was observed, $H\rightarrow 0$ extrapolation was too difficult
to prove the existence of the phase transition without doubt.
Lattice sizes used may be too small for this $\beta$.

\begin{figure}[h]
\centerline{\epsfxsize=14cm \epsfbox{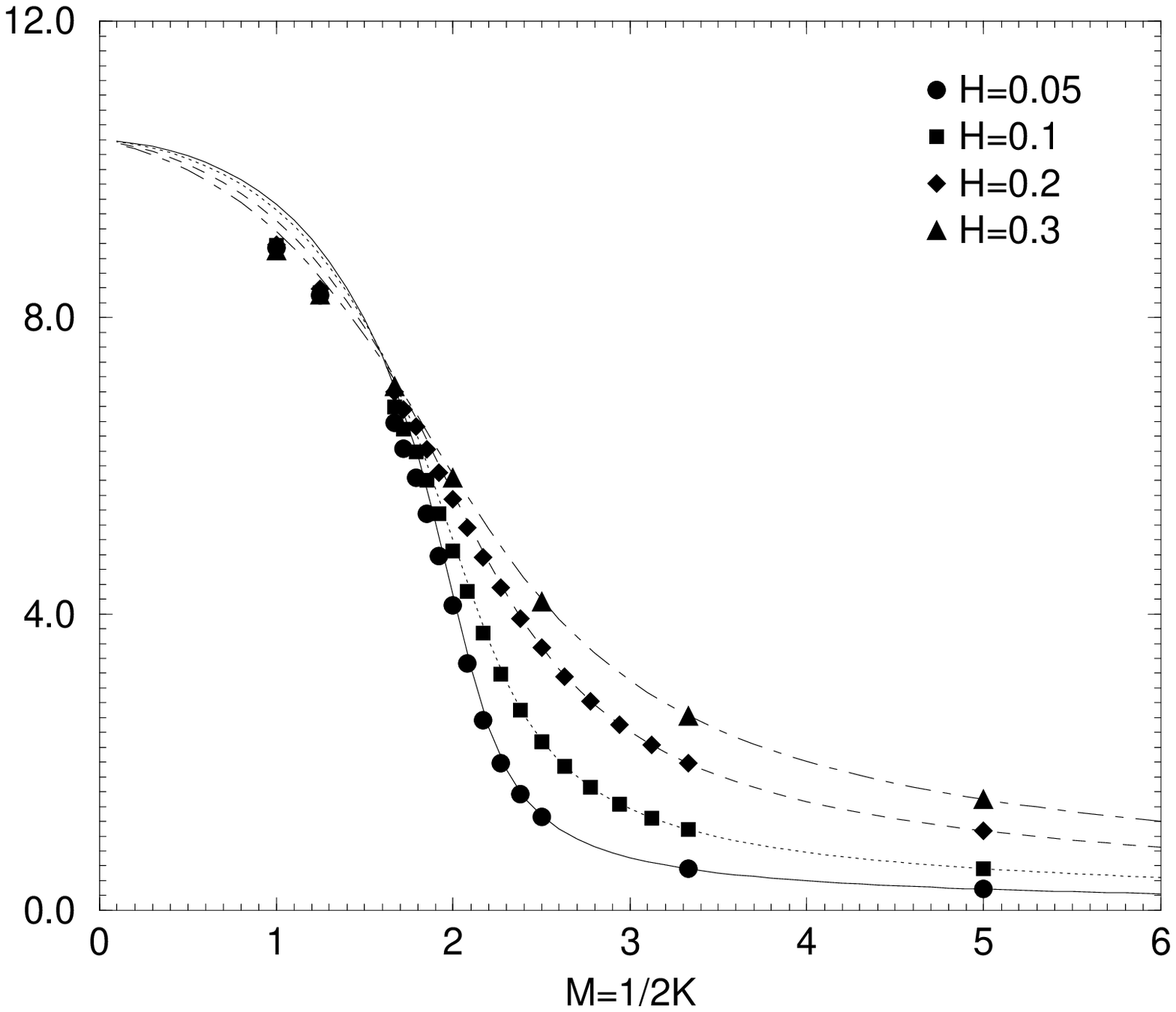}}
\caption{$\langle\bar\psi i\gamma_5\psi\tau^3\psi\rangle $
as a function of $M$
at $\beta =0$ on a $4^4$ lattice with $H=0.05,0.1,0.2,0.3$.
The lines are analytical prediction obtained at $\beta =0$ in the large
$N_c$ limit with the corresponding $H$.}
\label{ordb0}
\end{figure}

Since the numerical evidence of the phase transition is established for
the quenched case but is ambiguous for the case of the full QCD,
we carry out new full QCD simulation with $N_f =2$ at $\beta = 0$,
where the analytic expressions for the order parameters are available
for non-zero external field $H$.
In Fig.~\ref{ordb0}, $\langle \bar\psi i\gamma_5\tau^3 \psi \rangle$
is plotted as a function of $M$
at $\beta =0$ on a $4^4$ lattice with $H= 0.05,0.1,0.2,0.3$,
together with the analytic expressions with the corresponding $H$
in the strong coupling and large $N_c$ limits,
Since the agreement between the numerical data and the analytic predictions
are very good, it can be established at least in the strong coupling limit
that the parity and flavor symmetry breaking phase transition really occurs
at $M=M_c=2$ ($K=K_c=1/4$) and $\langle \bar\psi i\gamma_5\tau^3 \psi \rangle$
becomes non-zero in the $H\rightarrow 0$ extrapolation at $M<M_c$ ($K>K_c$).
We have checked that the finite size effect to the order parameter
is very small by comparing the data on a $4^4$ lattice with the data on
a $6^4$ lattice or with the analytic expressions corresponding to
the values in the infinite volume limit.

\begin{figure}[h]
\centerline{\epsfxsize=14cm \epsfbox{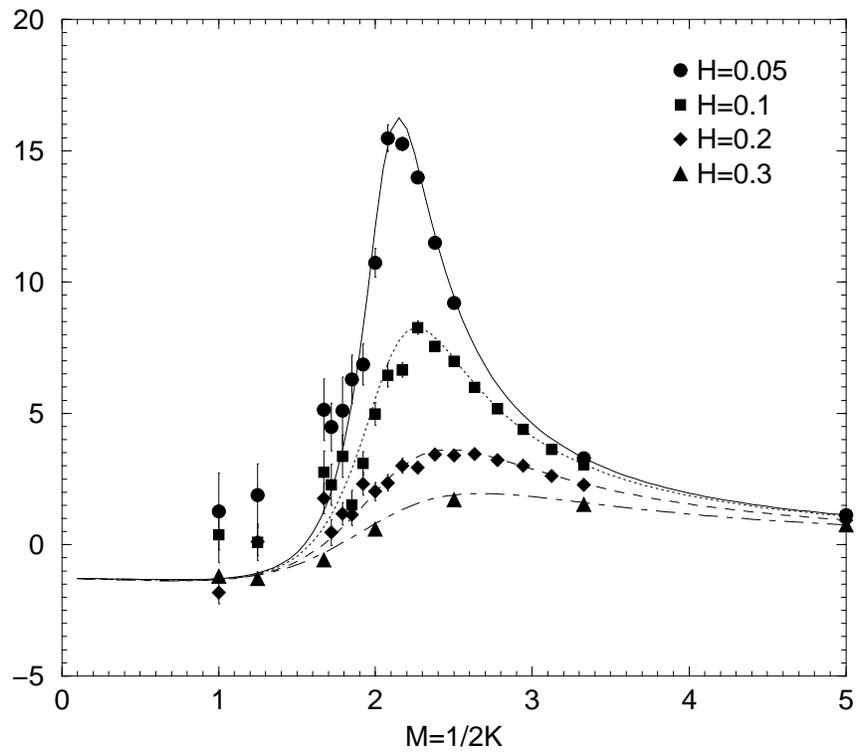}}
\caption{Susceptibility $\chi_{55}$
as a function of $M$ at $\beta =0$ on a $4^4$ lattice
with $H=0.05,0.1,0.2,0.3$.
The lines are analytical prediction obtained at $\beta =0$ in the large
$N_c$ limit with the corresponding $H$.}
\label{susp}
\end{figure}

As a further check we have calculated susceptibilities $\chi_{55}$,
$\chi_{00}$, and $\chi_{50}$, which are defined by
\begin{eqnarray}
\chi_{55} &=& \frac{1}{V}\langle [\sum_x \bar\psi i\gamma_5\tau^3 \psi (x)]^2
\rangle \\
\chi_{00} &=& \frac{1}{V}\langle [\sum_x \bar\psi \psi (x)]^2 \rangle \\
\chi_{50} &=& \frac{1}{V}\langle [\sum_x \bar\psi i\gamma_5\tau^3 \psi (x)]
\times [\sum_y \bar\psi \psi (y)] \rangle ,
\end{eqnarray}
both analytically and numerically.
For example, $\chi_{55}$ is plotted in Fig.~\ref{susp}.
The agreement between them are reasonably good, except some deviations
are observed deep in the broken phase.
Therefore,
the result for the susceptibilities also support our conclusion about
the existence of the phase transition.

If we consider the above result from the opposite side
it can be said that the success of our analytic calculations, except
the deviation in the susceptibilities deep in the broken phase,
establishes the validity of the strong coupling and large $N_c$
expansions near the limits.
Furthermore,
since the expansions predict the existence of the phase transition,
we can conclude that the parity and flavor symmetry breaking
is shown to exist in the strong coupling region.

Finally let me comment on the deviations in the susceptibilities
observed deep in the broken phase.
We think that the deviations are
caused by the massless Nambu-Goldstone charged pions appeared in the broken
phase, whose effect to the susceptibilities can not be included in
the analytic calculation of the large $N_c$ limit.
We expect the agreement will become better if we include the effect
in our analytic expressions.

\section{Phase structure in the weak coupling region}

In the previous section we have established the phase structure
in the strong coupling region. There are two critical lines
$M_c(\beta)$ and $ -M_c(\beta)$ in $\beta$ - $M$ plane,
which separate the broken phase from the symmetric phases:
$\langle \bar\psi i\gamma_5\tau^3 \psi \rangle \not= 0$
at $-M_c(\beta) < M < M_c(\beta)$ and
$\langle \bar\psi i\gamma_5\tau^3 \psi \rangle  = 0$ at
$M < -M_c(\beta) $ or $M_c(\beta) < M$.
In this section we investigate how this phase structure will be modified
in the weak coupling region.

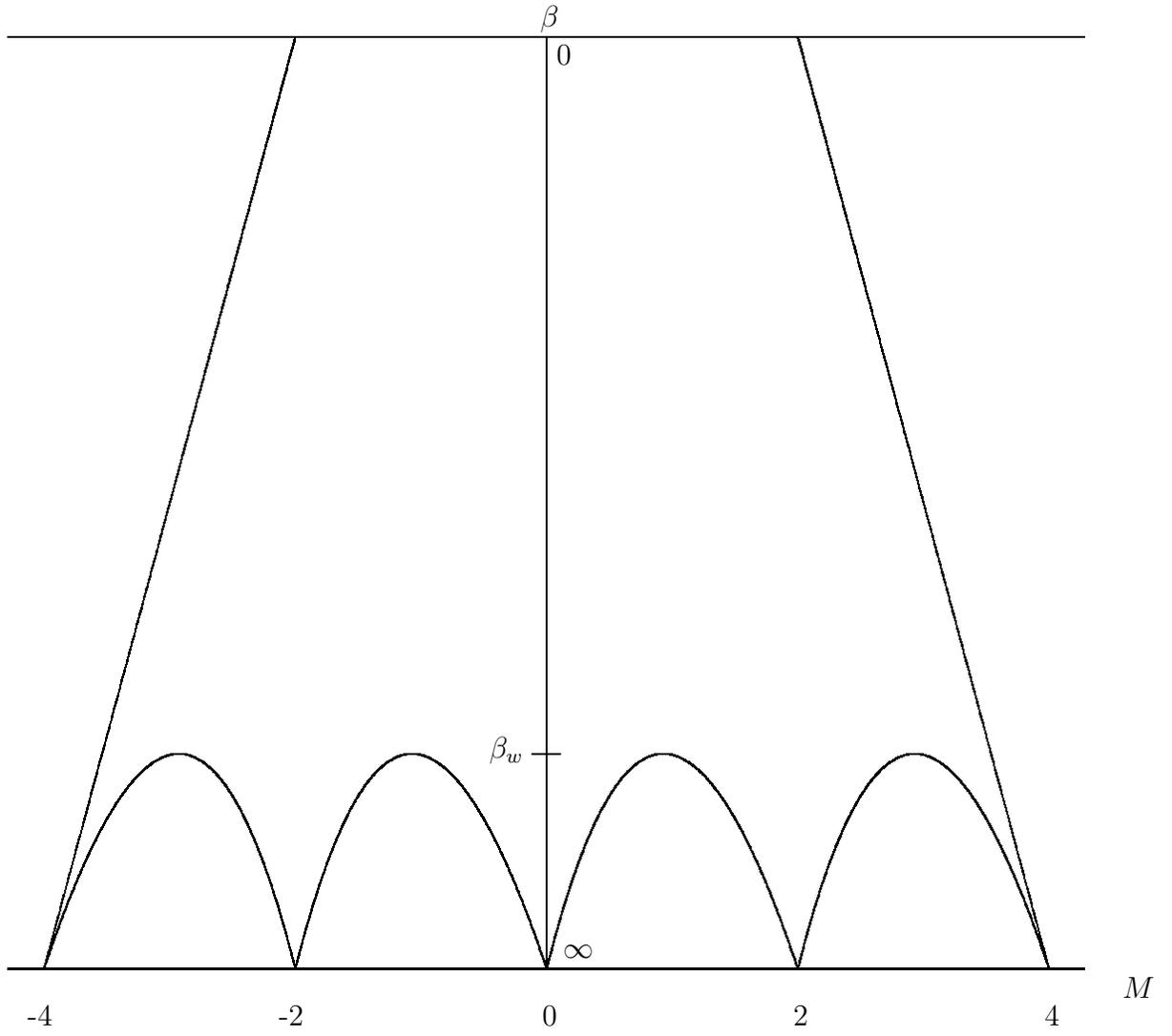
\begin{figure}[h]

\begin{center}

\setlength{\unitlength}{1mm}
\begin{picture}(170,150)
\put(10,10){\line(1,0){150}}
\put(85,10){\line(0,1){130}}
\put(10,140){\line(1,0){150}}
\put(87,10){\makebox(5,5){$\infty$}}
\put(165,5){\makebox(5,5){$M$}}
\bezier{1000}(15,10)(50,140)(50,140)
\bezier{1000}(15,10)(35,70)(50,10)
\bezier{1000}(50,10)(65,70)(85,10)
\bezier{1000}(155,10)(120,140)(120,140)
\bezier{1000}(155,10)(135,70)(120,10)
\bezier{1000}(120,10)(100,70)(85,10)
\put(12,1){\makebox(5,5){-4}}
\put(47,1){\makebox(5,5){-2}}
\put(83,1){\makebox(5,5){0}}
\put(118,1){\makebox(5,5){2}}
\put(153,1){\makebox(5,5){4}}
\put(83,140){\makebox(5,5){$\beta$}}
\put(85,135){\makebox(5,5){0}}
\put(83,40){\line(1,0){4}}
\put(77,38){\makebox(5,5){$\beta_w$}}
\end{picture}

\end{center}

\caption{The expected phase structure of lattice QCD with Wilson fermion in
$\beta$ - $M$ plane.}
\label{phase}
\end{figure}

In Fig.~\ref{phase} we propose the expected phase structure
of lattice QCD with 2 flavor Wilson fermions in the $\beta$ - $M$ plane.
At $\beta < \beta_w$ there are only two critical lines, while
there are ten critical lines at $\beta > \beta_w$ and 5
points where two lines meat at $\beta =\infty$. Each point
corresponds to one continuum limit, whose low energy spectra
are composed of a part of sixteen zero modes of the lattice fermion:
one mode near $p=(0,0,0,0)$ appears at $M=4$,
four modes near $p=(\pi,0,0,0)$ and its permutation at $M=2$,
six modes near $p=(\pi,\pi,0,0)$ and its permutation at $M=0$,
four modes near $p=(\pi,\pi,\pi,0)$ its permutation at $M=-2$, and
one mode near $p=(\pi,\pi,\pi,\pi)$ at $M=-4$.
Note that the phase diagram is shown to be symmetric under $M\rightarrow -M$
transformation.
It is also noted that the width of the broken phase in the weak coupling
should behave like $ C(g^2) \exp [ - B/g^2 ] $ where $B$ is some constant
and $C(g^2)$ is some regular function of $g^2$, since
the appearance of the width has not been detected
by the usual perturbative expansion,
and therefore should be non-perturbative.

There are several evidences for the existence of the multiple critical lines
in the weak coupling region. First of all, the phase structure
similar to Fig.~\ref{phase} has been found for the Gross-Nevue model
in 2 dimensions in the large $N$ limit, regulated on the lattice using
Wilson fermions\cite{aoki1,aoki3,EN}.
The existence of the multiple structure of the critical line
has been also suggested from
studies of eigen-value spectra of the Wilson fermion matrix
on quenched configurations\cite{SDB} and full QCD configurations\cite{BLLS}.

\begin{figure}[h]
\centerline{\epsfxsize=14cm \epsfbox{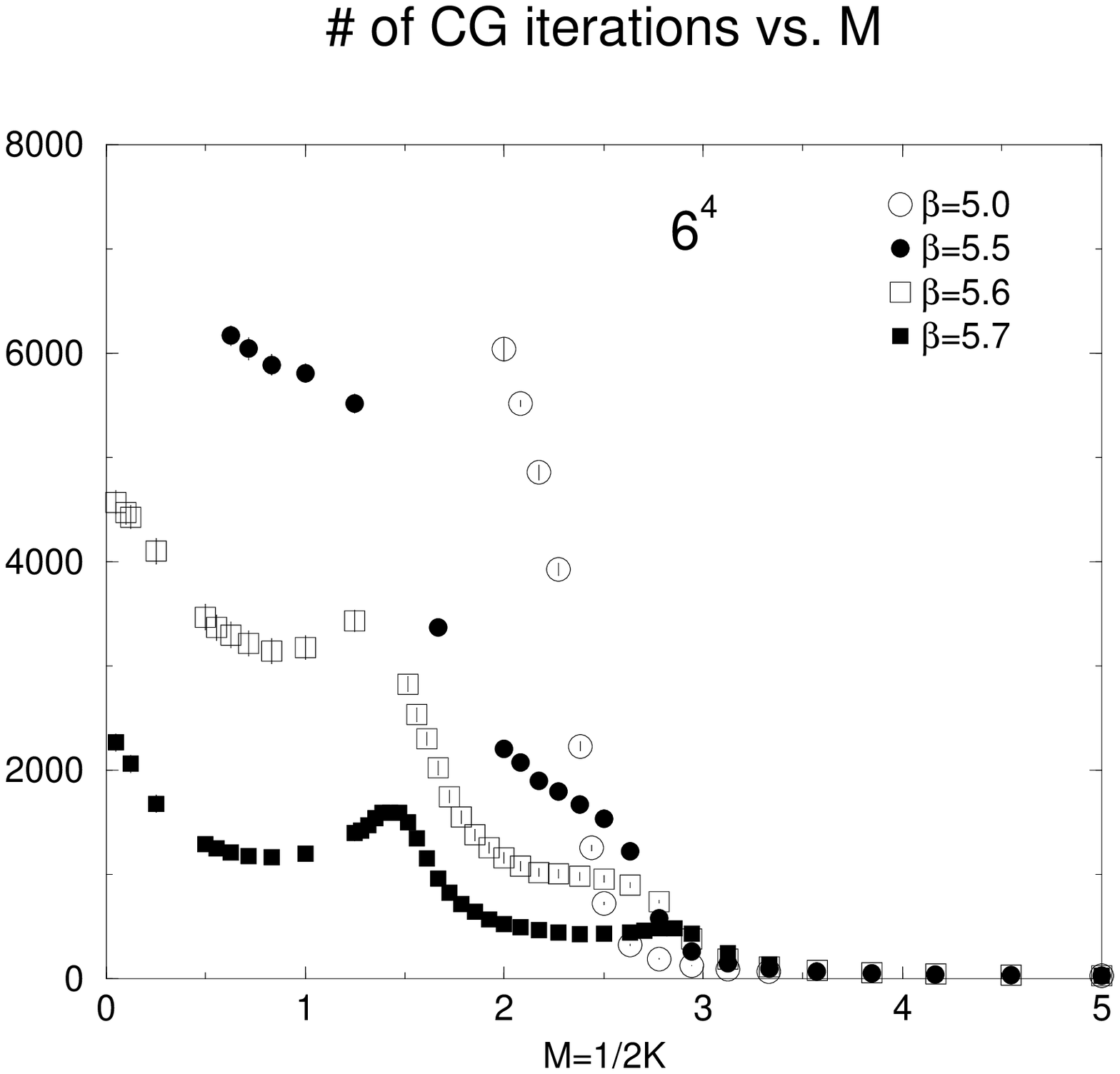}}
\caption{The number of CG iterations
as a function of $M$ at $\beta =5.0$, 5.5, 5.6 and 5.7
on $6^4$ lattices.}
\label{CG1}
\end{figure}

\begin{figure}[h]
\centerline{\epsfxsize=14cm \epsfbox{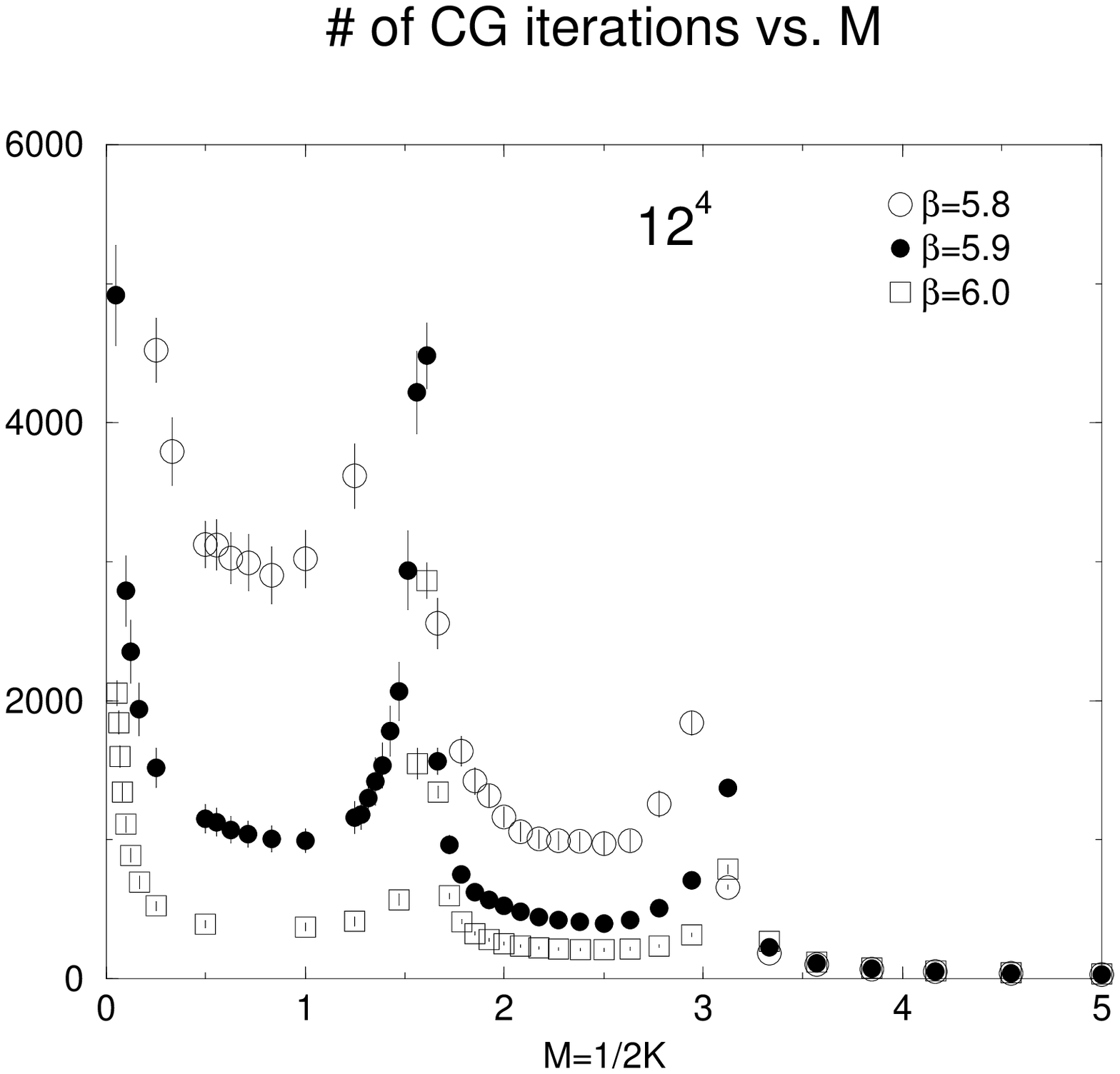}}
\caption{The number of CG iterations
as a function of $M$ at $\beta =5.8$, 5.9 and 6.0
on $12^4$ lattices.}
\label{CG2}
\end{figure}

To establish the multiple structure of the critical lines
we have performed new simulation using quenched configurations
{\it without} adding the external field $ i H \bar\psi \gamma_5\psi$.
We first monitored the quantity so-called ``pion norm'' :
$
\frac{1}{V}  \langle [\sum_x\bar\psi i\gamma_5\psi (x)]^2
\rangle $ .
This quantity becomes very large in the broken phase, due
to the contribution from zero modes of the Wilson fermion matrix.
Although some signals for the multiple structure of the broken phases
are observed in this quantity,
more clear indicator for the multiple structure of the broken phases
is found to be
the number of the CG iteration necessary to invert the Wilson fermion
matrix. This quantity becomes very large and
is proportional to the volume in the presence of zero modes
of the Wilson fermion matrix, which are expected to exist in the broken
phase.
We have plotted the number of CG iterations
as a function of $M$ at $\beta =5.0$, 5.5, 5.6, 5.7 on $6^4$ lattices
in Fig.\ref{CG1},
and at $\beta =$ 5.8, 5.9, 6.0 on $12^4$ lattices
in Fig.\ref{CG2}.
The multiple structure of the broken phase  is clearly observed
at $\beta \ge 5.6$.

\begin{figure}[h]
\centerline{\epsfxsize=14cm \epsfbox{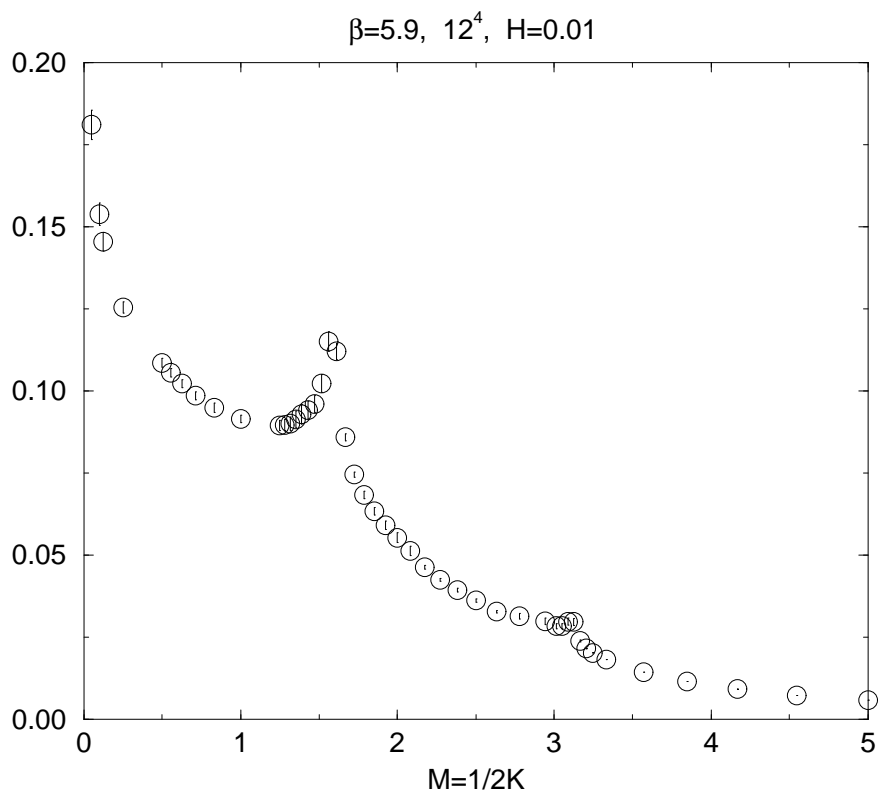}}
\caption{$\langle\bar\psi i\gamma_5\psi\rangle $
as a function of $M$ at $\beta =5.9$ on a $12^4$ lattice
with $H=0.01$.}
\label{ordw}
\end{figure}

Finally we calculate again the order parameter
$\langle \bar\psi i\gamma_5\psi \rangle$
with small non-zero external field, $H = 0.01$,
in the weak coupling region ($\beta =5.9$).
As seen from Fig.~\ref{ordw}, the multiple structure of the broken phases
is again observed

\section{Conclusion and further investigations}

The phase transition which causes or is caused by the massless pion
appeared in  lattice QCD with the Wilson fermion formulation
seems to exist at $K = K_c$.
In the strong coupling region at $ K > K_c$
it is established that
$\langle \bar\psi i\gamma_5\psi\rangle \not=0$ for the quenched
QCD and $\langle \bar\psi i\gamma_5\tau^3\psi\rangle \not=0$
for the 2 flavor full QCD.
The evidence of the multiple structure of the critical lines
are also given in the weak coupling region in the case of the
quenched QCD.

As mentioned in the introduction,
it is very important to study
the phase structure of lattice QCD at finite temperature.
In the case of the Wilson fermion formulation
one should understand, based on the finding of this report,
the relation between the critical line discussed in this report and
the finite temperature phase transition line.
Recently such an investigation has been carried out\cite{AUU}
for the 2 flavor QCD. The study concludes that
the critical line of the massless pion at finite temperature
forms a cusp at finite gauge coupling, and
that the line of thermal transition runs past the tip of the cusp without
touching the critical line. The detail can be found in ref.~\cite{AUU}.

\section*{Acknowledgements}

The part of this work has been completed in collaboration with
A. Ukawa and T. Umemura.
Numerical calculations for the present work have been
carried out at Center for Computational Physics, at University of Tsukuba.
This work is supported in part by the Grants-in-Aid of
the Ministry of Education(Nos. 04NP0701, 06740199).

\newpage



\vfill

\end{document}